\documentstyle[11pt,paspconf,epsf]{article}

\begin{document}

\title{RR Lyrae Variables in Globular Clusters and the
       Second-Parameter Phenomenon}

\author{M. Catelan and A. V. Sweigart}
\affil{Laboratory for Astronomy and Solar Physics, Code 681, NASA/GSFC,
       Greenbelt, MD 20771, USA}

\author{J. Borissova}
\affil{Institute of Astronomy, Bulgarian Academy of Sciences,\\
       72 Tsarigradsko chauss\`ee, BG-17884 Sofia, Bulgaria}

\begin{abstract}
We propose to use RR Lyrae variables in globular clusters (GC's)
to probe the origin of the second-parameter phenomenon.
\end{abstract}


\section*{1. Introduction}

Until recently, the view that age is the second parameter of 
horizontal-branch (HB) morphology has been widely accepted.
Recent developments suggest, however, a much more complicated
scheme where several parameters may be playing important
r\^oles (Stetson, VandenBerg, \& Bolte 1996; Ferraro et al.
1997; etc.).

Second-parameter candidates can be divided into two
categories: those which do {\it not} affect the HB luminosity
(age, mass loss on the red giant branch), and those which {\it do}
(envelope helium abundance $Y$, core mass at helium-flash). 
{\it We thus propose to use RR Lyrae variables, whose properties 
are strongly sensitive to the HB luminosity (e.g., van Albada \& 
Baker 1971), as a vertical diagnostic of the HB morphology}.

\section*{2. Approach to the Problem}

The determination of RR Lyrae ``equilibrium temperatures" 
($T_{\rm eq}$) relies on the transformation of mean
colors into temperatures (e.g., Sandage 1990; Carney, Storm, 
\& Jones 1992, CSJ92). CSJ92 demonstrated the 
existence of a tight relationship involving $T_{\rm eq}$, 
periods, blue amplitudes ($A_B$), and metallicity [Fe/H] for
field RR Lyraes. We have rederived the CSJ92 relationship
from data in their Table 4, but using the latest [Fe/H] values
from Layden (1994). Stars SS Leo and TT Lyn were excluded from
the fit, due to their poorly determined parameters.

To our surprise, we discovered that the period term in the
$T_{\rm eq}$--period--$A_B$--[Fe/H] relationship does {\it not} 
lead to a significant reduction in the residuals to justify its
inclusion. In fact, {\it determining temperatures from the
periods may easily mask the effects of luminosity variations
upon cluster-to-cluster period shifts, and should thus be
avoided whenever possible}. Our new relationship reads:

\begin{equation}
 \frac{5040}{T_{\rm eq}({\rm K})} = (0.850\pm 0.013) - 
    (0.073\pm 0.008) \, A_B - 
    (0.010\pm 0.003) \, {\rm [Fe/H]}.
\end{equation}
This relationship provides an excellent match to the 
CSJ92 temperatures. The rms deviation is only 56.9 K,
compared with 54 K from eq.~(16) in CSJ92 
(which does include a period term).

\section*{3. Applications to Globular Clusters}

Assuming eq.~(1) to be also valid for GC RR Lyraes,
we have computed temperatures and measured the 
``period shifts" $\Delta\log\,P$ at fixed $T_{\rm eq}$
with respect to the ``reference" cluster M3, for ab-type RR 
Lyraes in a set of 34 GC's. Blazhko variables were avoided. 
Our main findings may be summarized as follows: 
a) Clusters with similar metallicity may present 
significantly different $\Delta\log\,P$ values 
(e.g., NGC~6171 vs. NGC~6712); b) The bimodal-HB GC
NGC~1851 shows very large $\Delta\log\,P$'s (and also 
{\it very} high $A_B$'s), the opposite occuring in 
NGC~6229 (which also has a bimodal HB); c) Among the Oosterhoff 
type II (OoII) GC's, we find a range in $\Delta\log\,P$'s, 
with a mild correlation with HB type (which however can
be almost entirely ascribed to NGC~5986); d) In the mean, OoII
GC's do not show very large $\Delta\log\,P$'s; e) The RR Lyrae 
variables in the metal-rich clusters 47~Tuc and NGC~6388 have 
very large $\Delta\log\,P$'s; f) $\omega$~Cen variables have large 
$\Delta\log\,P$'s, but a detailed analysis of the correlation 
between $\Delta\log\,P$ and [Fe/H] in this cluster must await 
the resolution of the discrepant [Fe/H] distributions found for 
its giants and RR Lyraes (Suntzeff \& Kraft 1996).

\section*{4. A Caveat}
Possible helium abundance differences between field and GC
RR Lyraes (Sweigart 1997) may have to be further investigated, 
as far as the period-shift effect is concerned. In particular, from 
Marconi's (1997) models, we find that {\it M15 variables lie in 
the region of the $A_B - T_{\rm eq}$ plane most affected by $Y$, 
whereas the M3 ones fall mostly in the region not sensitive to $Y$}.

\newpage

\title {New Possible Variables in the Outer-Halo\\
        Globular Cluster Palomar 3}

\author{J. Borissova and N. Spassova}
\affil{Institute of Astronomy, Bulgarian Academy of Sciences,\\
       72 Tsadrigradsko chauss\`ee, BG 1784 Sofia, Bulgaria}

\author{M. Catelan}
\affil{NASA/GSFC, Code 681, Greenbelt, MD 20771, USA}

\author{V. D. Ivanov}
\affil{Steward Observatory, University of Arizona, Tucson, AZ 85721, USA}

\begin{abstract}

We present a list of ten possible variable stars in the globular
cluster Pal~3. Seven are new suspected variables. The variability
of the RR Lyr candidate reported by Burbidge \& Sandage (1958),
as well as of the Pop. II Cepheid and RR Lyr from Gratton \& 
Ortolani (1984), are confirmed.
\end{abstract}

\section*{1. Introduction}
The limited number of variability studies for the outer-halo globular
cluster Pal~3 (Burbidge \& Sandage 1958, BS58; Gratton \& Ortolani 1984,
GO84) has prompted us to undertake a new survey for short-period variables.
Special motivation for our study was provided by the possible presence of 
a Pop. II Cepheid in such a red-HB cluster (GO84). As well known, Pop. II
Cepheids are usually found in clusters with blue-HB tails only (e.g., Smith 
\& Wehlau 1985).

\section*{2. Observations and data reduction}

Our analysis was based on 35 $V$ and $I$ CCD frames obtained on 3 nights 
at the 1.54m telescope operated by Steward Observatory, University of Arizona, 
and at the 2m telescope of NAO ``Rozhen" (Bulgaria). The photometric reductions
were carried out using the {\sc DAOPHOT} package available in {\sc IRAF}. The 
instrumental magnitudes were transformed to the standard $VI$ system.

\section*{3. Results}

Using the variability search techniques of Welch \& Stetson (1993) and
Kadla \& Gerashchenko (1982), comparison of the brightness and
night-to-night variations of the stars, and light curve 
analysis, we reached the following conclusions:
\begin{enumerate}
\item We confirm the variability of the RR Lyr star candidate from BS58;
\item We confirm the variability of the Pop. II Cepheid and the RR Lyr
      star (No.~185) suspected by GO84;
\item We find seven  new variable star candidates (cf. Table 1);
\item Stars V1, V2, V9 and V10 show variability according to all  
      variability criteria, and are thus very likely to be true variable 
      stars. V1, V2 and V10 are probable ab-type RR Lyr stars (according to 
      the light curve analysis), while V9 is probably a red variable;
\item Stars V3, V4 and V5 satisfied only some of the variability criteria, 
      and their status remains uncertain.
\end{enumerate}
Unfortunately, our small number of CCD frames and irregular time intervals 
did not permit us to obtain representative light curves and periods.
The limited data allow only to estimate a preliminary period $P = 0.47$~days 
for V2.

The x and y coordinates in Sawyer-Hogg's (1973) catalog system and the 
estimated mean $V$ magnitudes of candidate variable stars are given in Table~1.

\begin{table}[h]
\caption {Mean $V$ magnitudes for candidate variable stars in Pal~3}
\begin{tabular} {lrrccc}
\hline\\

             &            &       &   &  JD 244+    &      \\
            &            &       &50465.59    &50467.53 & 50490.54     \\
 Name &   x        &   y   & $V$&  $V$  & $V$  \\
            &  (arcsec)  &(arcsec)&   &      &      \\
\hline\\
V1          &   $-3.2$   &  $-2.3$  &21.06  &20.38 &20.27  \\
V2          &   $15.6$   &   $4.1$  &20.22  &20.62 &20.73  \\
V3          &  $-57.7$   &  $14.0$  &20.26  &20.56 &20.34  \\
V4          &   $25.6$   &  $18.6$  &20.45  &20.44 &20.61  \\
V5          &  $-60.7$   &  $35.0$  &20.48  &20.48 &20.62  \\
BS58        &  $-29.2$   &  $-4.8$  &20.25  &20.66 &20.25  \\
No. 155     &   $-0.8$   &   $4.4$  &20.82  &20.71 &20.62  \\
Pop. II Cep.&   $18.6$   &  $17.0$  &19.26  &19.41 &19.44  \\
V9          &   $-8.2$   &   $0.3$  &18.63  &18.70 &18.72  \\
V10         &  $-42.4$   &  $-5.9$  &20.13  &20.54 &20.44  \\
\hline
\end{tabular}
\label{Tab01}
\end{table}

\newpage

\title{On the Production of Bright RR Lyrae Variables in Metal-Rich
       Globular Clusters}

\author{A. V. Sweigart and M. Catelan}
\affil{NASA/GSFC, Code 681, Greenbelt, MD 20771, USA}


\section*{1. Second Parameter Effect in NGC~6388 and NGC~6441}
Recent HST observations have revealed that the metal-rich 
globular clusters (GC's) NGC~6388 and NGC~6441 contain an unexpected population
of hot horizontal-branch (HB) stars and therefore exhibit the well-known 
second parameter effect (Rich et al. 1997). Most surprisingly,
the HB's of these clusters have a pronounced upward slope with decreasing 
$B-V$, with the mean HB luminosity at the top of the blue tail being 
roughly 0.5~mag brighter in $V$ than the well-populated red clump, which 
itself also slopes upward. Differential reddening is not the cause of 
these sloped HB's (Piotto et al. 1997).

The second parameter effect has often been attributed to differences
in age or mass loss on the red-giant branch (RGB). However, canonical
simulations show that increasing the assumed age or
RGB mass loss will move an HB star blueward, but will {\it not} 
increase a star's luminosity (Fig. 1a). This indicates that the blue HB
population in NGC~6388 and NGC~6441 does not arise from either an
older cluster age or greater mass loss. Something else must be affecting 
the HB morphology in these two clusters. {\it Nature may therefore 
be giving us an important clue for understanding the origin of the 
second parameter effect.}

We have begun an extensive theoretical 
study to determine the cause
of the sloped HB's in NGC~6388 and NGC~6441. Three scenarios are being 
studied:

\begin{itemize}
\item A high cluster helium abundance scenario, where the HB 
      morphology is determined by long blue evolutionary loops;
\item A rotation scenario, where the core mass in the HB models
      is increased by internal rotation during the RGB phase;
\item A helium-mixing scenario, where deep mixing on the RGB
      enhances the envelope helium abundance (Sweigart 1997).
\end{itemize}

\begin{figure}[t]
\plotfiddle{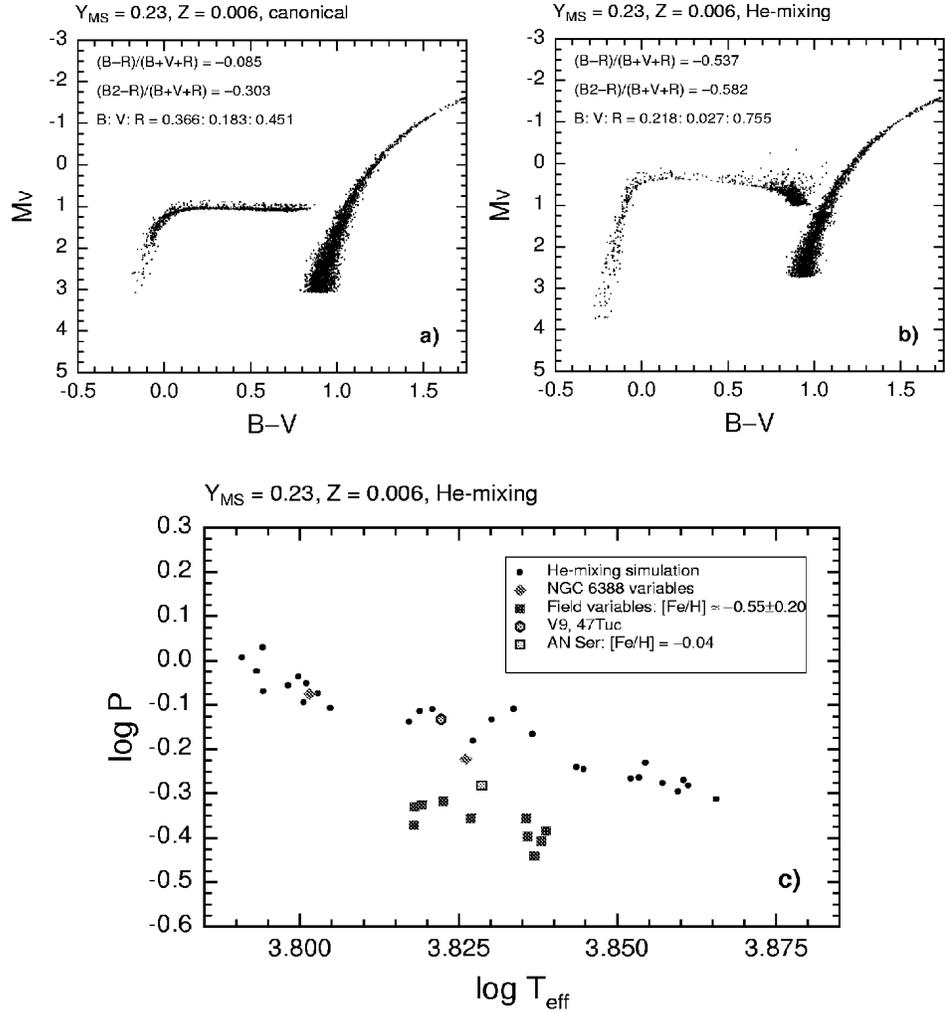}{4.4in}{0}{102.5}{102.5}{-315}{-357}
\caption{``Canonical" HB simulation (panel a), compared to a
  non-canonical one with an HB slope similar to that
  in NGC 6388 and NGC 6441 (panel b). The pulsation properties
  for this non-canonical case are compared to the 
  observational data in panel c. Temperatures have been derived
  as described by Catelan, Sweigart, \& Borissova (1998).}
\end{figure}  

We have found that all three of these scenarios predict sloped
HB's (e.g., Fig. 1b) with anomalously bright RR Lyrae variables. 
This prediction is consistent with the long pulsation periods of 
the two known RRab Lyrae variables in NGC~6388 (Silbermann et al. 
1994), as demonstrated in Fig.~1c. If confirmed, these scenarios
would have important implications for stellar evolution and for 
the use of HB stars as standard candles to determine GC distances
and ages.

Our full poster paper can be found at {\tt astro-ph/9708174}.

\acknowledgments
A. V. S. was supported by NASA grant NAG5--3028.
This work was performed while M. C. held a National Research
Council--NASA/ GSFC Research Associateship.

\end{document}